\documentclass{aa}
\usepackage{graphicx}
\newcommand{\HI}{{\ion{H}{i}}}
\newcommand{\kms}{km s$^{-1}$}
\newcommand{\miriad}{{\texttt {MIRIAD}}}
\newcommand{\ltsima} {$\; \buildrel < \over \sim \;$}

\begin{document}
\title{\HI\ Absorption in High-Frequency Peaker Galaxies}
\subtitle{   }

\author{
M. Orienti\inst{1,2} \and
R. Morganti\inst{3,4} \and
D. Dallacasa\inst{1,2} 
}
\offprints{M. Orienti}
\institute{
Dipartimento di Astronomia, Universit\`a di Bologna, via Ranzani 1,
I-40127, Bologna, Italy \and 
Istituto di Radioastronomia -- INAF, via Gobetti 101, I-40129, Bologna,
Italy \and
Netherlands Foundations for Research in Astronomy, Postbus 2, 7990 AA
Dwingeloo, The Netherlands \and
Kapteyn Astronomical Institute, University of Groningen, PO Box 800,
9700 AV Groningen, The Netherlands 
}
\date{Received \today; accepted ?}

\abstract{
{WSRT observations have been used to investigate the presence of neutral
hydrogen in extremely young radio galaxies. These objects were selected from a
sample of High-Frequency Peakers (HFPs). We detect 2 of the 6 observed galaxies
confirming previous detection of \HI\ in these objects. 
In the case of \object{OQ~208} -
for which discrepant results were available - 
we confirm the presence of a broad
($\sim$ 1800 km s$^{-1}$), blue-shifted and shallow \HI\ absorption. No
significant changes in the \HI\ profile have been found between the two epochs
of the observations. The intriguing result is that the derived
\HI\ column densities and upper limits obtained for the most compact
sources, do not follow the inverse correlation between the column
density and the linear size found for CSS/GPS sources. This can be explained -
assuming the gas is already in a torus/disk structure - by a combination of
the orientation and the extreme compactness of the sources. 
In this scenario, our line of sight to the source would 
intersect the torus in its inner region with low
optical depth due to high spin and kinetic temperatures. There is no evidence,
with the exception of \object{OQ~208}, 
of unsettled, high column density gas still
enshrouding the young radio sources. This can be due to the low filling factor
of such a medium.  }

{}
\keywords{
galaxies: active -- galaxies: evolution -- radio continuum:
galaxies -- radio lines:galaxies}
          
}     
\titlerunning{\HI\ Absorption in High-Frequency Peaker Galaxies}
\maketitle
\section{Introduction}

A substantial amount of gas and dust is often found in the central
regions of galaxies harbouring an Active Galactic Nucleus (AGN).  This gas
provides important information on the physical conditions and processes in
these near-nuclear regions (Morganti et al. \cite{rm04a}; Peck et
al. \cite{pt99}). A component of this gas, the atomic hydrogen, can be
studied in radio-loud objects via absorption detected against the strong
continuum source. Since years, a large number of detailed studies has
been done (see e.g.  Heckman et al. \cite{hec83}; van Gorkom et
al. \cite{vg89}; Morganti et al. \cite{rm01},
\cite{rm04a}, \cite{rm05}; Vermeulen et al. \cite{rv03}; Pihlstr\"{o}m et
al. \cite{yp03}) and they have shown
that the neutral hydrogen can be associated with different structures, from
circumnuclear tori to fast outflows.  These results have been crucial
for constructing our present picture of the nuclear structure 
of radio loud AGN.

The medium enshrouding the nuclear regions also plays an important role
in the growth and evolution of the radio source (Labiano et
al. \cite{al05}).  The presence of significant amount of gas in
  young radio
galaxies is supported by a larger incidence of \HI\ absorption in these
objects (Vermeulen et al. \cite{rv03}; Pihlstr\"{o}m et al. \cite{yp03})
compared to what typically found in old and larger radio sources (van Gorkom
et al. \cite{vg89}; Morganti et al. \cite{rm01}). This dense medium
is likely the result of the merger that triggered the radio source
(Morganti et al. \cite{rm04}). Thus, in the initial phase
of the radio source, the radio jet likely has to make its way through a
relatively dense medium that surrounds the active nucleus. 

The radio galaxies classified as ``Compact-Steep Spectrum'' and ``GHz-Peaked
Spectrum'' are believed to represent the early stages of the radio
activity (Fanti et al. \cite{cf95}; Readhead et al. \cite{rh96}; Snellen et
al.  \cite{sn00}).  These radio sources are intrinsically small (sub-galactic
size, i.e.  $<$ 15 kpc) and bright (P$_{\rm 1.4 GHz}$ $>$ 10$^{25}$ W/Hz),
characterised by a convex radio spectrum, peaking at frequencies between
100 MHz and a few GHz. Their radio morphologies appear to be the
scaled-down version of powerful edge-brightened radio galaxies, with luminous
mini-lobes ($\sim$ 0.1 up to few kpc) and weaker jets and cores.  Both
kinematic (Polatidis \& Conway \cite{PC03}), and spectral (Murgia \cite{Mu03})
studies strongly support the youth scenario, indicating ages of 10$^{3}$ --
10$^{5}$ years. 

Given the relatively high detection rate of \HI\ absorption in these objects
(Vermeulen et al. \cite{rv03}; Pihlstr\"{o}m et al. \cite{yp03}), it is of
particular interest to investigate the characteristics of the nuclear
interstellar medium (ISM) in even younger radio sources. In the youth
scenario, the anti-correlation between the turnover frequency and the linear
size (O'Dea \cite{odea98}), which is indicative of the age, suggests
that the youngest sources have the highest turnover frequency.  Therefore, the
``High-Frequency Peaker'' (HFP) radio sources, characterised by the same
properties of CSS/GPS, but with the spectral turnover
occurring at frequencies higher than 5 GHz, are
good candidates to be {\it newly born} radio sources, with ages of about
10$^{2}$--10$^{3}$ years (Dallacasa
\cite{dd03}).

This paper reports on the result of observations searching for \HI\
absorption in a sample of 6 HFP radio sources, selected from the
Dallacasa et al.  (\cite{hfp0}) bright HFP sample, and suitable to be
observed at the Westerbork Synthesis Radio Telescope (WSRT).
\begin{table*}
\begin{center}
\begin{tabular}{ccccccccccccc}
\hline
\hline
&&&&&&&&&&&&\\
Source    &Other   &LS&$z_{\rm opt}$&$ \nu_{\rm obs}$ &
Resol.&r.m.s.&S$_{\rm obs}$&S$_{\rm HI}$&$\tau_{\rm peak}$&$\Delta$v&Log(N$_{\rm
    HI}$)&$z_{\rm HI,peak}$\\
J2000     &name    &pc& &MHz&km s$^{-1}$&mJy/b/ch &mJy&mJy& &km s$^{-1}$& & \\
(1)       &(2)  & (3) &(4)&(5)&(6)&(7)&(8)&(9)&(10)&(11)&(12)&(13)\\
\hline
& & & & & & & & & & & &  \\
\object{J0003+2129}&        &22&0.452&977.96&4.3&6.1&50&$<$12.2&$<$0.15& &$<$21.43& \\
\object{J0111+3906}&\object{OC\,314}&22$^{a}$&0.668&851.32&6.9&2.8&170&52.7&0.44&100&21.90&0.6687 \\
\object{J0655+4100}&        &$<$1&0.02156&1390.03&4.3&0.7&239&$<$1.4&$<$0.006& &$<$20.0& \\
\object{J1407+2827}&\object{OQ\,208} &10&0.773&1318.60&4.5&1.1&826&5&0.005&1800&20.9&0.0769\\
\object{J1511+0518}&        &7&0.084&1309.96&9.5&1.1&80&$<$2.2&$<$0.02& &$<$20.6&\\
\object{J1623+6624}&        &$<$1&0.203&1180.38&4.5&0.7&129&$<$1.4&$<$0.01& &$<$20.3&\\
&&&&&&&&&&&&\\
\hline
\end{tabular} 
\vspace{0.5cm}
\end{center}    
\caption{Physical and observational parameters of the 6 HFP galaxies
  observed with WSRT. Columns 1, 2: source names; 
  Column 3: projected linear sizes (Orienti et al. 2006); Column 4:
  optical redshift; Column 5: central frequency; Column 6: channel
  resolution; Column 7: 1$\sigma$ noise level in  the line cube;
  Column 8: continuum flux density taken from our WSRT data at the
  observed frequency; Column 9: peak flux
  density of the absorption line, measured on the spectral image; Column 10:
  optical depth; Column 11: the width of the \HI\ absorption line: for
  \object{J0111+3906} the FWHM is given, in the case of \object{OQ\,208} we give the FWZI, due to
  the complexity of the line profile. Column 12: \HI\ column density derived
  from 
  $N_{\rm \HI} = 1.82 \times 10^{18} T_{\rm spin} \tau_{\rm peak} \Delta V $
  cm$^{-2}$, a T$_{\rm spin}$ of 100 K has been assumed; Column 13:
  the redshift of the peak \HI\ absorption. 
  The line flux density, the optical depth and the \HI\ column
  density upper limits have been computed assuming the 2$\sigma$ noise level,
  a line width of 100 km s$^{-1}$ and T$_{\rm spin}$ of 100 K, as in Vermeulen
  et al. (\cite{rv03}).
  $a$: For the source \object{J0111+3906}, the projected linear
  size is taken from Owsianik et al. (\cite{ocp98}).} 
\label{taboss}
\end{table*}  

\section{The sample}

The bright HFP sample (Dallacasa et al. \cite{hfp0}) has been constructed by
cross-correlating the Green Bank Survey (87GB) at 4.9 GHz and the NRAO VLA Sky
Survey (NVSS) at 1.4 GHz. Only the sources with a flux density brighter than
300 mJy at 4.9 GHz and with a rising spectrum steeper than $\alpha =
- 0.5$ (S$\propto \nu^{-\alpha}$; 
i.e. to avoid turnover frequencies $\leq$ 5 GHz) 
have been selected as HFP candidates. To tackle the contamination of variable sources, like
blazars, all the candidates have been simultaneously observed with the VLA at
eight different frequencies from 1.4 to 22 GHz. The final sample consists of 55 sources ($\sim$
3\% of the sources exceeding 300 mJy in the 87GB): 11 galaxies with redshifts
ranging from 0.02 to 0.67; 33 quasars with higher redshifts, typically between
0.9 and 3.5; 5 BL Lac objects, while 6 sources still lack optical identification
(Dallacasa et al. \cite{dfs}; Dallacasa et al., in preparation).  
Since the selection of these sources is based on their
  simultaneous radio spectra at a single epoch, there is still some
  contamination by beamed radio sources, whose emission is temporarily
  dominated by a self-absorbed jet component.
Further
multi-frequency VLA observations to exclude the long-term variability (Tinti et
al. \cite{st05}) and pc-scale resolution morphological analysis (Orienti et
al. 2006), have been carried out to confirm genuine and young HFPs. 

As shown by Orienti et al. (\cite{mo06}), there is a clear segregation
in radio morphology between galaxies and quasars. The majority of
galaxies (9/11) show a Compact-Symmetric-Object(CSO)-like morphology,
while 28/33 quasars display ``Core-Jet'' or unresolved structures. 
This is consistent with the idea that the HFP spectrum in galaxies and
quasars originates in different regions: mini-lobes and/or hot-spots
in galaxies, compact regions related to the core and the jet base in
quasars, in agreement with what found in GPS (Stanghellini et
al. \cite{cs05}) and bright CSS samples (Dallacasa et al. \cite{dd95};
Fanti et al. \cite{rf90}).\\
Since the radio sources hosted in galaxies likely represent a population of young radio sources, six out of the eleven HFP galaxies,
have been observed to search for \HI\ absorption.
The selected objects have
recessional velocities such that the redshifted \HI\ line is shifted to
frequencies that are known to be relatively free of radio interference
(RFI) at the WSRT. Their redshift ranges from 0.02 and 0.67, and the flux densities at
1.4 GHz exceed 90 mJy. 

For one of the five remaining HFP galaxies, WSRT observations were precluded,
since a well known RFI (GSM) affects the frequency band, while for the other
four sources, the optical redshift is not available.

\section{WSRT observations and data reduction}

The WSRT observations have been carried out in different runs from 29 January
to 12 September 2005 in dual orthogonal polarisation mode.  Two sources in the
sample were already known to have \HI\ absorption: from Carilli et
al. (\cite{cari98})
in the case of \object{J0111+3906} and Vermeulen et al. (\cite{rv03}) and 
Morganti et al. (\cite{rm05}) in the case of \object{OQ~208}.  
In these two sources, the goal of
the new observations was to investigate the possible presence of variability
in the \HI\ absorption and/or a broad component in
their absorption spectrum. In particular, in the case of \object{OQ~208}, the two
previous observations report different characteristics (i.e. width) of the
absorption. 

Of the six observed galaxies, those with $z < 0.3$ have been observed for 6
hours each, using the L-band receiver, with 1024 spectral channels covering a
20 MHz wide observing band. The sources with redshift higher than 0.4 have
been observed for 12 hours each, using the UHF-high band, with 1024 spectral
channels. The 20 MHz band, allowed to cover a velocity range around the
velocity centroid of approximately $\pm$ 2200 \kms\ at $z=0.07$, and
approximately $\pm$ 3500 \kms\ at z=0.6.  Only the source
\object{J0003+2129} has been
observed with a 10 MHz wide band ($\sim$ $\pm$ 1500 \kms), since known RFI
would have affected a wider band.  Table \ref{taboss} summarizes the
observational and physical parameters of the sources.

The data reduction was carried out using the \miriad\ package.  The quasars
\object{3C\,286}, \object{3C\,48} and \object{3C\,147} were observed to calibrate the absolute flux
density scale and the bandpass.  Data were inspected for time-limited and
baseline specific interference, and bad data were removed before solving for
the gain- and band-pass calibrations.  The two orthogonal linear polarisations
were added together to improve the signal-to-noise ratio of the final
spectrum.  The continuum subtraction was done by using a linear fit through
the line-free channels (i.e. all the channels outside a range of $\pm$ 1000
\kms\ from the centroid frequency) of each visibility record and subtracting
this fit from all the frequency channels (using the \miriad\ task
``UVLIN''). Particular attention has been given to the case of \object{OQ~208}
where shallow absorption has been detected (see below).
After several trials, the final continuum subtraction was obtained using as
line-free those channels outside the range $-1700/+600$ from the centroid
frequency. 
The baselines of the spectra
obtained after the continuum subtraction result flat, indicating that no broad
and shallow line features have been included.

After the continuum subtraction, line cubes were produced using uniform
weighting, Hanning smoothed and inspected. For the two sources where \HI\
absorption was detected (\object{J0111+3906} and \object{OQ~208}), the spectra obtained at the
location of the peak of the radio continuum are shown in Figs. 1 and 2.  For
every source, the r.m.s. noise was estimated from the line cube and the values
are given in Table \ref{taboss}.

In order to make a reliable comparison between our values and what found in
CSS/GPS samples, the upper limits to the optical depth and the \HI\ column
density have been computed following Vermeulen et al. (\cite{rv03}),
i.e. assuming the 2$\sigma$ noise level, a line width of 100 km s$^{-1}$ and
T$_{\rm spin}$ of 100 K (see Tab. \ref{taboss}).

Finally, we have used the line-free channels to produce a continuum image of
each target. The continuum emission is always unresolved at the
resolution of the WSRT (typically about $20^{\prime\prime}$), with the only
exception of \object{J0111+3906}, known to possess a relatively weak component
on such angular scale (Baum et al. \cite{sb90}). The
continuum emission derived from our data is slightly
resolved to the East, consistent with the large-scale emission seen by Baum et
al. (\cite{sb90}).

\section{Results}

Of the six galaxies observed, we detect \HI\ absorption only in the two
sources (\object{J0111+3906} and \object{OQ\,208}) 
in which the presence of absorption was
already reported by previous studies. In addition to this, we find \HI\ in
emission in two companion galaxies of \object{J0655+4100}.  We discuss below these
three objects in more detail.

\subsection{The source \object{J0111+3906}}

The radio source \object{J0111+3906} 
is associated with a narrow emission line galaxy
at $z = 0.668$ (Carilli et al. \cite{cari98}).  The source
is classified as a Compact Symmetric Object (CSO), due to its VLBI-scale
morphology (Taylor et al. \cite{gt96}), but 
it also shows an extended emission on
the kpc-scale, interpreted in terms of recurrency of the radio activity (Baum
et al. \cite{sb90}).

\begin{figure}
\begin{center}
\includegraphics{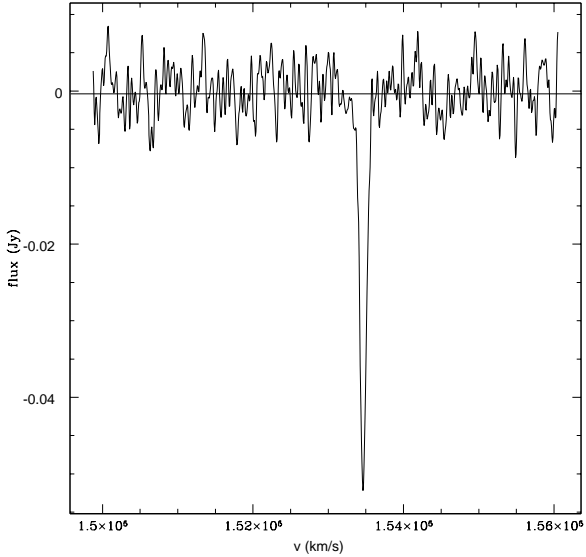}
\vspace{8.5cm}
\caption{The \HI\ absorption profile detected in \object{J0111+3906}.
The velocity is in the optical heliocentric convention.}
\label{j0111}
\end{center}
\end{figure}

\HI\ absorption was already reported by
Carilli et al. (\cite{cari98}), based on narrow-band WSRT observations. We
detect a strong \HI\ absorption line peaking at $\nu_{\rm \HI,peak}= 851.03$
MHz ($z_{\rm \HI,peak} = 0.6687$, see Fig. \ref{j0111}). The \HI\ absorption
is only detected against the peak of the continuum, coincident, therefore,
with the VLBI-scale structure. The line width is FWHM = 100 km s$^{-1}$ and the
optical depth $\tau_{\rm peak} = 0.437$.  With these values, assuming a spin
temperature of 100 K, we obtain an \HI\ column density of 8.05 $\times$
10$^{21}$ cm$^{-2}$. These values are in good agreement with those found by
Carilli et al. (\cite{cari98}).\\ 
No broad absorption (i.e. with FWHM $>$ 100 km s$^{-1}$), 
with optical depth higher than $\tau_{2\sigma} \sim 0.03$, has
been detected.

\begin{figure} 
\begin{center}
\includegraphics{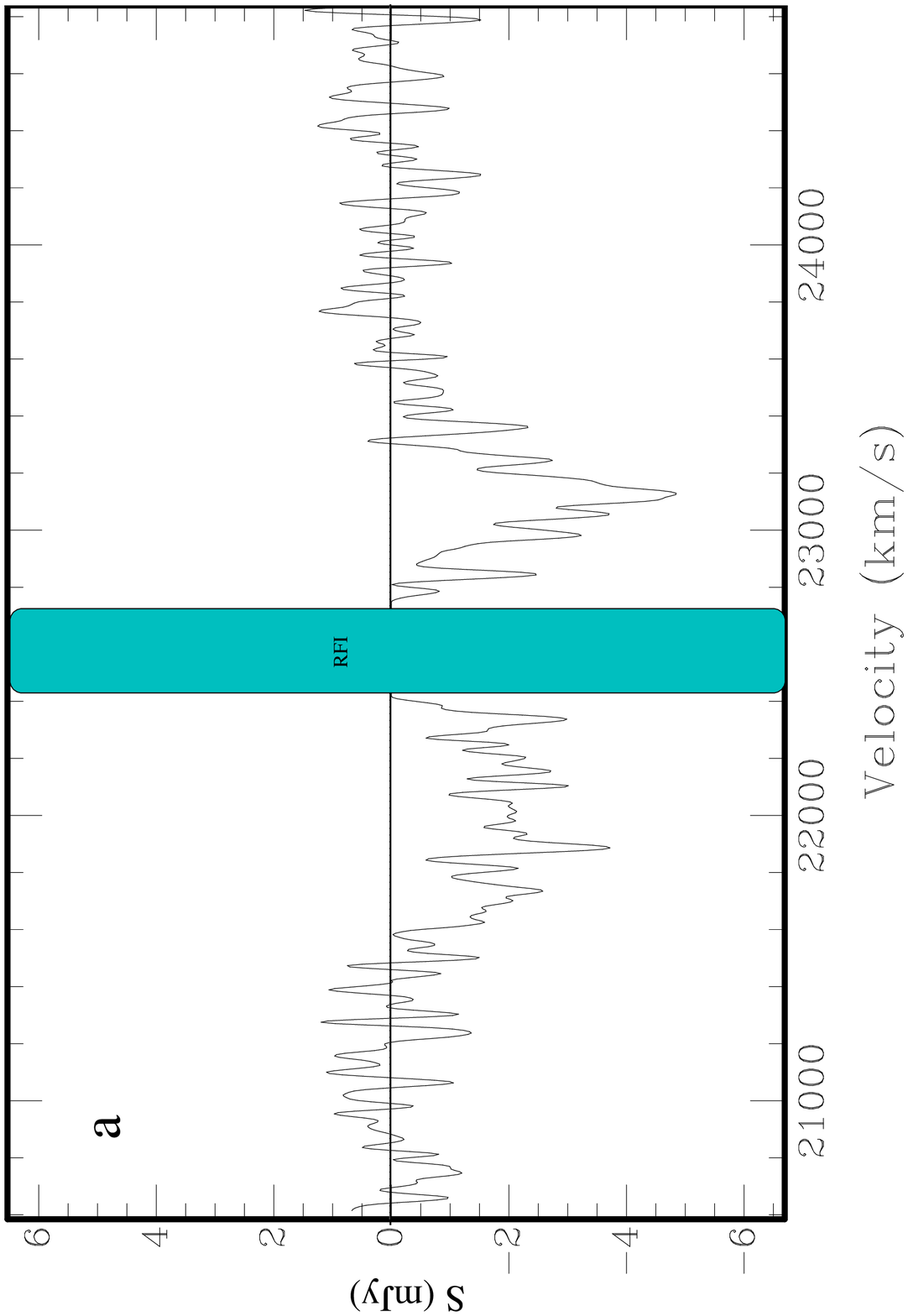}
\includegraphics{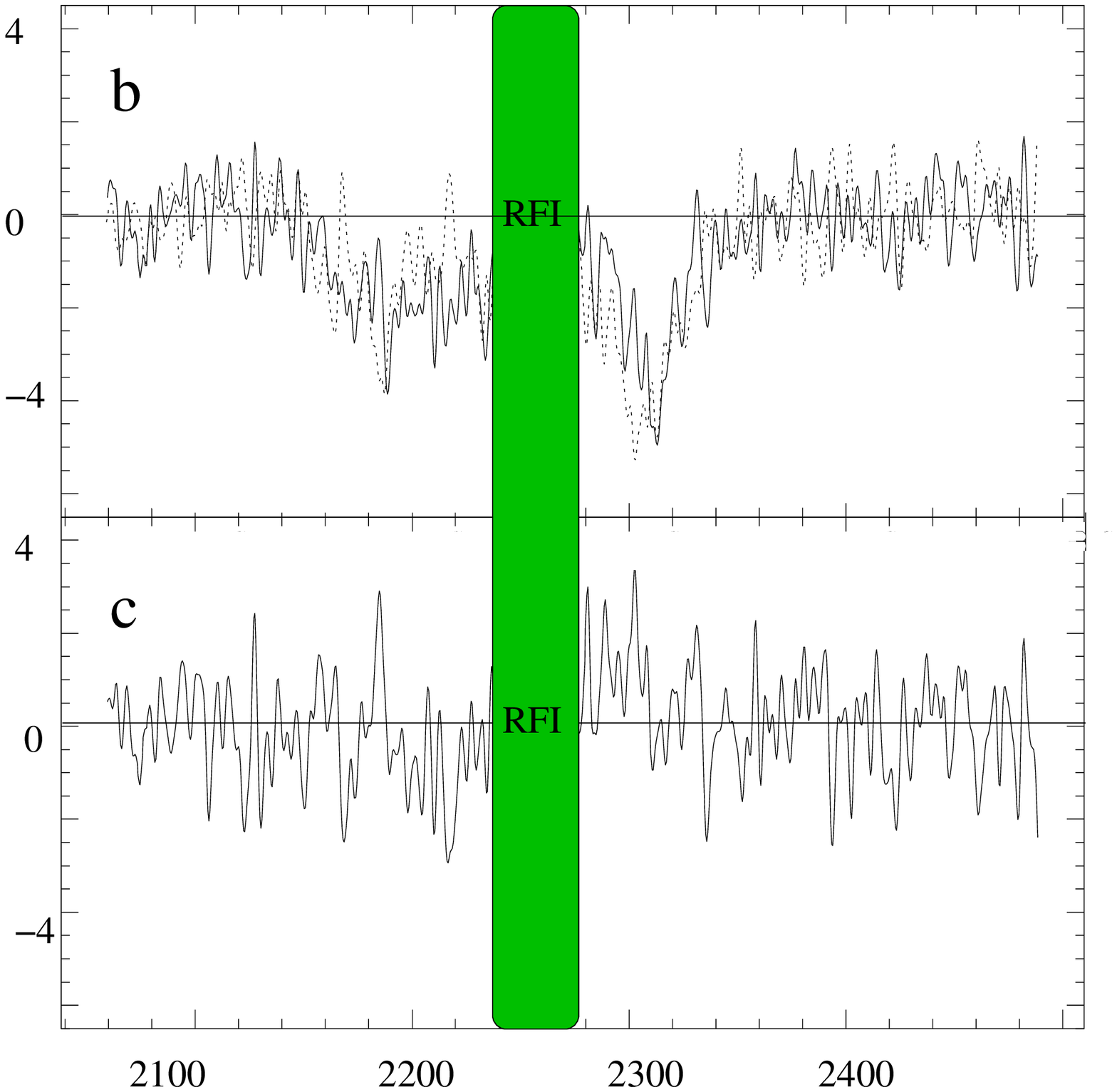}
\vspace{15.5cm}
\caption{{\it a:} The \HI\ absorption profile detected in \object{OQ\,208}
  from our new
  WSRT observations. The velocity is in the optical heliocentric
  convention. Strong RFI is present in the line profile. 
  {\it b:} The two-epoch line profiles of \object{OQ\,208} one
  superimposed to the other; the continuous line represents data from
  Morganti et al. (\cite{rm05}), while the dotted line represents our new
  data. {\it c:} the residuals obtained after subtracting the
  two-epoch line profiles.}
\label{oq208}
\end{center}
\end{figure}

\subsection{The source \object{OQ\,208}}

In the radio source \object{OQ~208} we detect a shallow ($\tau_{\rm peak}$ $=$
0.005) and very broad (FWZI $\sim$ 1800 \kms) \HI\ absorption
(Fig. \ref{oq208}a). This absorption extends from $\sim$ 21650 to $\sim$ 23450
\kms, therefore mostly blue-shifted compared to the systemic velocity
22957 \kms\ from Gallimore et al. (\cite{gjf99}). The line profile has a peak
located at $\nu_{\rm \HI,peak} = 1318.6$ MHz ($z_{\rm \HI,peak} = 0.0773$).  Assuming a
T$_{\rm spin}$ of 100 K, we obtain an \HI\ column density of 8.0 $\times$
10$^{20}$ cm$^{-2}$.

The \HI\ absorption in \object{OQ~208} was already detected by both Vermeulen et
al. (\cite{rv03}) and Morganti et al. (\cite{rm05}) although with different
results. Vermeulen et al. (\cite{rv03}) found an \HI\ absorption centered on
the same frequency ($\nu = 1318.9$ MHz) as in our observations but with a
width of 256 \kms only. The likely reason for the discrepancy with our results
is the fact that the maximum available bandwidth at the WSRT at the
time of the survey of Vermeulen et al. (\cite{rv03}) was not large enough to
allow the detection of such broad features.  More recently, the presence of a
broad \HI\ absorption in \object{OQ~208} was reported by Morganti et al. (\cite{rm05})
and interpreted as a fast outflow of neutral hydrogen. Fig. \ref{oq208}b shows
the comparison between the \HI\ profile derived from the observations
presented here and what was found by Morganti et al. 
(\cite{rm05}). In Fig. \ref{oq208}c, the residuals obtained after subtracting
the profiles taken in two different epochs are shown. The figure shows that
there is no significant evidence of changes in the
\HI\ profile corresponding to an optical depth upper limit of  
$\tau$ $\sim$ 0.002.

It is worth mentioning that the radio source \object{OQ\,208} 
is associated with a
broad emission line galaxy at z=0.07658.  The radio structure is characterised
by two compact lobes, separated by about 10 pc, highly asymmetric in terms of
flux density ratio (Stanghellini et al. \cite{cs97}).  In this source, the
shape of the radio spectrum could either arise from synchrotron
self-absorption or free-free absorption by an external absorber embedding the
lobes (Kameno et al. \cite{k00}).  Furthermore, X-ray observations (Guainazzi
et al. \cite{guai04}) have led to the discovery of a Compton-thick obscured
AGN where the column density of the absorber is 10$^{21}$ cm$^{-2}$.  As
suggested by Guainazzi et al. (2004), in this source the jets are likely
piercing their way through this Compton-thick medium which is pervading the
nuclear environment.  The outflow detected in \HI\ would be another indication
of this process.

\subsection{The source \object{J0655+4100}}

\begin{figure} 
\begin{center}
\includegraphics{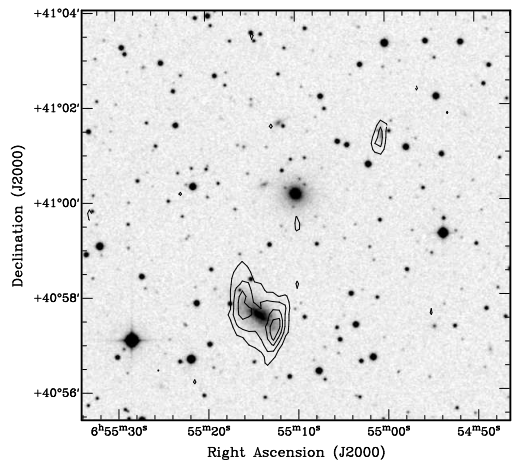}
\vspace{7.0cm}
\caption{\HI\ total intensity contours superimposed onto a DSS optical image
centered on the radio galaxy \object{J0655+4100}. Two nearby
  galaxies are clearly detected. Contours levels are: 2.35, 3.50, 5.68
  and 7.51 $\times$ 10$^{20}$ atoms cm$^{-2}$.} 
\label{group}
\end{center}
\end{figure}

Although no evidence of neutral hydrogen gas has been found in the
radio galaxy \object{J0655+4100}, 
an elliptical galaxy at $z = 0.02156$ (V$_{\rm sys} =
6464$ \kms; Marcha et al.  \cite{mm96}), we find instead, \HI\ in emission in
two nearby galaxies.  
One companion is the spiral galaxy \object{UGC03593} where, 
centered at 6694 \kms,
$\sim 2\times10^9$ M$_{\odot}$ of \HI\ have been detected.
This galaxy is located 65 kpc South-East of \object{J0655+4100}.
The second companion displays a systemic velocity of 6334
\kms\ and is located to the North-West at 55 kpc from the HFP source.  
In the Digitized Sky Survey (DSS) image it is associated with a faint galaxy
for which we find an \HI\ mass of $\sim$ 2$\times$ 10$^{8}$ M$_{\odot}$.  This
galaxy has not been previously catalogued\footnote{Based on results from the
NASA/IPAC Extragalactic Database (NED).}.  The two galaxies form a
physical compact group with \object{J0655+4100} (Fig. \ref{group}) supporting the idea
that young radio sources reside in groups, as suggested by other works
(Stanghellini et al. \cite{cs93}), based on the excess of galaxy density in
the optical images of GPS radio sources.\\

\section{Discussion}

The detection in \HI\ absorption of only two of the six observed
galaxies is somewhat surprising.  Analysing a sample of 41 CSS/GPS radio
sources, Pihlstr\"{o}m et al. (\cite{yp03}) have found an inverse correlation
between the source linear size and the \HI\ column density: smaller sources
($<0.5$ kpc) have larger \HI\ column density than the larger
sources ($>0.5$ kpc).  
Following this correlation, one would have expected very high N$_{\rm
HI}$ in the smallest sources, like HFPs. 

\begin{figure} 
\begin{center}
\includegraphics{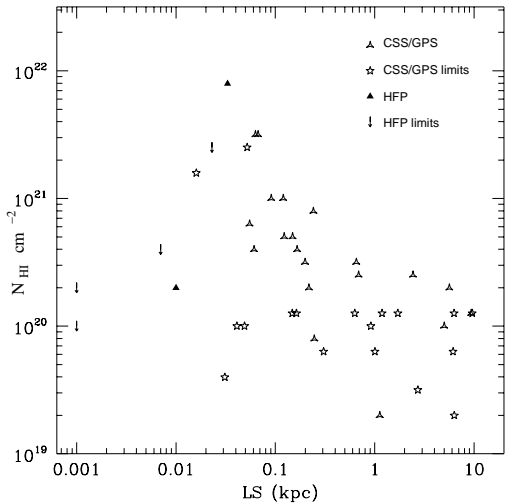}
\vspace{7.0cm}
\caption{Absorbed \HI\ column density versus projected linear sizes. The
HFP galaxies do not seem to follow the trend found for CSS/GPS
sources. The CSS/GPS values have been taken from Pihlstr\"{o}m et al.
(\cite{yp03}).}
\label{nhi}
\end{center}
\end{figure}

In Fig.  \ref{nhi} we show the inverse correlation found by Pihlstr\"om
et al. (\cite{yp03}) with the addition of the
values obtained from the sources presented in this paper.
The sensitivity reached by the new observations, enables us
to set tight upper limits to the N$_{\rm HI}$ for the sources with no \HI\
detection.  Interestingly, in our sample, 
the smallest sources (with size \ltsima 10 pc) 
appear to deviate from the extrapolation of the trend
found by Pihlstr\"om et al. (\cite{yp03}). Given the limited number of
objects, we cannot establish whether for very compact radio sources a break
down of the correlation occurs.  
If we consider that 15 out of 41 sources of Pihlstr\"{o}m et
al. (\cite{yp03}) sample have \HI\ column density below
2$\times$10$^{20}$ cm$^{-2}$, the probability to find two sources randomly chosen
both below such limit is $\sim$ 13\%. This means that the small
column density is possibly a feature typical of extremely young radio
galaxies.  
We also point out that the correlation from
Pihlstr\"om et al. (\cite{yp03}) has a large scatter, and upper limits to the
column density are present at all linear sizes.  
With all this in mind, it is interesting to consider the possible
origin of the lack of \HI\ absorption found in \object{J0655+4100} 
and \object{J1623+5524}.

The trend observed by Pihlstr\"{o}m et al.  (\cite{yp03}) has been explained
with both a spherical and axi-symmetric gas distribution, with a radial power
law density profile, as well as a disk model.  Indeed, a torus-like
distribution of neutral gas has been found to be consistent with the
observations in a number of GPS sources (i.e.  \object{4C\,31.04} by 
Conway\cite{c96}; \object{1946+708} by Peck et al.  \cite{pt99}).  
In this scenario, the absence of high
\HI\ column density in very compact sources can be explained by both
the orientation and the extreme compactness of the sources.  Within the
framework of the AGN unification scheme, the central engine is surrounded by a
disk of ionised gas and shielded by an obscuring torus of atomic and molecular
gas (i.e.  Urry \& Padovani \cite{up95}).  Evidence of ionised disks with
radii between a few tens and one hundred pc has been claimed in
this class of sources (Kameno et al.  \cite{k00}), as responsible of free-free
absorption in the optically thick region of the radio spectrum.

\begin{figure} 
\begin{center}
\includegraphics{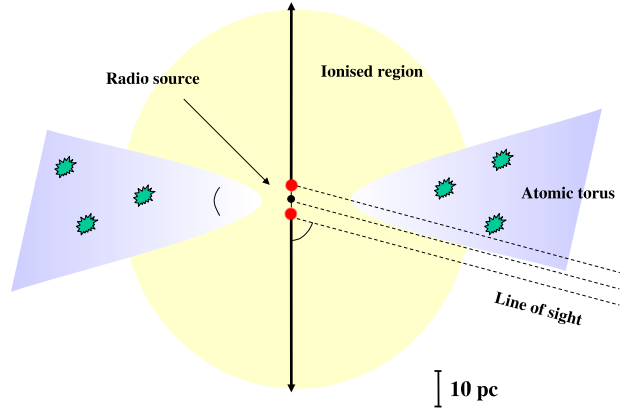}
\vspace{7.5cm}
\caption{Cartoon of a possible orientation of the HFP sources and
  circumnuclear torus. The toroidal structure is perpendicular to the
  source axis. The torus model is dependent on the opening angle
  $\theta$ and the viewing angle $\alpha$. The scale shown is
  approximate: the inner radius of the torus and the sizes of the
  denser clumps are not known.}
\label{model}
\end{center}
\end{figure}

The fundamental ingredient to trace \HI\ absorption is the presence of
background flux density.  Given the extremely small linear sizes of the
HFP sources (see Fig.  4 and Tab.  \ref{taboss}), the line of sight likely passes
mainly through the ionised region, piercing the torus only in its very inner
part, if we are looking at the receding lobe.  
This could be the case if the torus axis is aligned to the radio jet, and its
half-opening angle is, for example, $\theta$ $\sim$ 20$^{\circ}$ (as illustrated
in Fig.  \ref{model}), consistent with the constraint $\theta$ $\leq$
30$^{\circ} - 40^{\circ}$ by Granato et al.  (\cite{gdf97}), and 
an orientation of the line
of sight between 45$^{\circ}$ and 70$^{\circ}$ (with respect to the radio jet).

It should be noted that the gas closer to central engine is hotter, both in
terms of kinetic and spin temperature, therefore a longer path length would
not necessary imply a higher optical depth.  Moreover, recent work by
Hatziminaoglou et al. (\cite{hatz06}) focused on the properties of the dusty
tori in AGN, has shown the possibility of low optical depth tori.  Therefore,
in this scenario, it is possible that for these sources of a few pc in
size (or even less), we are missing the \HI\ absorption due to both
sensitivity limitations and orientation effects.

As described in Sect. 1, in addition to the presence of 
circumnuclear tori,
in such young radio sources one
could have expected to detect also the
presence of unsettled gas, i.e. gas not necessarily located in a circumnuclear
torus or disk-like structure but still enshrouding the radio source. We do not
find  evidence for this.  Again, this could be partly due to the very
small emitting area of the background radio source, that decreases the
probability to detect the absorption.  Clouds indicating the presence of a
rich and clumpy interstellar medium in the centre of the radio source have
been found as in the case of \object{4C~12.50} (Morganti et
al. \cite{rm04}), \object{3C\,49}
and \object{3C\,286.3} (Labiano et al. \cite{al05}).  
In \object{4C\,12.50} a cloud of $\sim
20\times 60$ pc has been detected: such a cloud would have covered most of the
HFPs in our sample. Thus, our results are possibly indicating that the filling
factor of such clouds is low, as it could be the case in \object{4C~12.50}.

Finally, the great majority of the fast outflows detected in both
  pc- and kpc-scale radio galaxies
have an \HI\ optical depth of only $\tau < 0.006$ (Morganti et
al. \cite{rm05}). Our observations are not sensitive enough to 
reach such low values, therefore the
lack of detection of broad absorption (except for the case of
  \object{OQ~208} where the
continuum is indeed strong enough) indicates that \HI\ outflows are
characterized by low optical depths.

\section{Conclusions}

An \HI\ absorption search has been carried out with the WSRT for a
sample of 6  HFP radio galaxies. We confirm the detection of
\HI\ in absorption in 2 galaxies.  One source, \object{J0111+3906}, 
is characterised by a line
width of $\sim$ 100 \kms\ and a high optical depth of $\tau = 0.44$. In the
other source (\object{OQ~208}), the line profile is very broad ($\sim$ 1800 \kms),
blue-shifted and shallow, with a maximum optical depth of $\tau = 0.005$. In
the remaining 4 galaxies no evidence of \HI\ absorption has been detected.

Although this result does not seem to follow the inverse correlation
found (Pilhstr\"{o}m et al. \cite{yp03}) between the column density and the
linear sizes, it can be explained by orientation effects in a torus scenario,
in which our line of sight intersects the torus in its inner region where the
low optical depth is due to high spin and kinetic temperature. Since these 4
sources have faint flux densities, optical depths $\leq$ 0.01 are not
detectable due to sensitivity limitations. Therefore, the \HI\
absorption in an object with the same spectral characteristics of
\object{OQ\,208} but
with a fainter flux density, cannot be detected by our observations.

Although HFP sources do not seem to follow the correlation between
\HI\ column density and linear size found for CSS/GPS sources, 
this does not imply that we are looking at a different class of objects,
instead of the youngest tail of a radio source population. Our results
suggest that on linear scales smaller than few tens of parsecs, the \HI\
column density is much lower than one would have expected on the basis of 
the work of Pilhstr\"{o}m et al. (\cite{yp03}).  As a consequence, compact
and rather faint (due to self-absorption) sources, such as these HFPs, are not
the most suitable class of objects to investigate the \HI\ absorption on such
a small scale.

\begin{acknowledgements}
We are grateful to  T.A. Oosterloo for helpful suggestions on the
best use of the software \miriad.
MO acknowledges the Kapteyn Astronomical Institute of the University
of Groningen and ASTRON for their hospitality during this project. 
Part of this research was funded by RadioNet and the Nova-Marie Curie
Fellowship programme. We thank the referee, Ren\'e Vermeulen, 
for useful comments which improved the paper.
The WSRT is operated by ASTRON (The Netherlands Foundation for
Research in Astronomy) with support from the Netherlands Foundation
for Scientific Research (NWO). This research has made use of the
NASA/IPAC Extragalactic Database (NED), which is operated by the Jet
Propulsion Laboratory, California Institute of Technology, under
contract with the National Aeronautics and Space Administration.\\
\end{acknowledgements}

\end{document}